\documentclass[showpacs,preprintnumbers,amsmath,amssymb,twocolumn,superscriptaddress,prb]{revtex4}
\usepackage{graphicx}
\usepackage{amsfonts}
\usepackage{amsmath, amsthm, amssymb}
\usepackage{dsfont}
\usepackage{times}
\usepackage{subfigure}

\newcommand{\bq}{\begin{equation}}
\newcommand{\eq}{\end{equation}}

\newcommand{\hz}{\hat{\mathbf{z}}}

\def\i{\mathrm{i}}

\begin{document}
\title[Tunneling conductance and local density of states
 in time-reversal symmetry breaking superconductors under the influence of an external magnetic field]
{Tunneling conductance and local density of states
 in time-reversal symmetry breaking superconductors under the influence of an external magnetic field}

\author{Mihail A. Silaev}
\affiliation{Institute for Physics of Microstructures, Russian Academy of Sciences, 603950 Nizhny Novgorod GSP-105, Russia}
\author{Takehito Yokoyama}
\affiliation{Department of Applied Physics, Nagoya University, Nagoya, 464-8603, Japan and CREST, Nagoya, 464-8603, Japan}
\author{Jacob Linder}
\affiliation{Department of Physics, Norwegian University of
Science and Technology, N-7491 Trondheim, Norway}
\author{Yukio Tanaka}
\affiliation{Department of Applied Physics, Nagoya University, Nagoya, 464-8603, Japan and CREST, Nagoya, 464-8603, Japan}
\author{Asle Sudb\o}
\affiliation{Department of Physics, Norwegian University of
Science and Technology, N-7491 Trondheim, Norway}

\date{Received \today}
\begin{abstract}
\noindent We consider different effects that arise when time-reversal symmetry breaking
superconductors are subjected to an external magnetic field, thus rendering the superconductor to be
in the mixed state. We focus in particular on two time-reversal symmetry breaking order parameters which
are believed to be realized in actual materials: $p+\i p'$-wave and $d+\i s$- or $d+\i d'$-wave. The first order
parameter is relevant for Sr$_2$RuO$_4$, while the latter order parameters have been suggested to exist
near surfaces in some of the high-$T_c$ cuprates. We investigate the interplay between surface states
and vortex states in the presence of an external magnetic field and their influence on both the tunneling
conductance and the local density of states. Our findings may be helpful to experimentally identify the symmetry of
unconventional time-reversal symmetry breaking superconducting states.
\end{abstract}

\maketitle

\section{Introduction}\label{sec:introduction}

Recently, considerable attention has been devoted to the chiral superconducting phase which is believed to be realized
 in the $p$-wave triplet superconductor\cite{Pwave} Sr$_2$RuO$_4$. The chiral state of a $p$-wave
 superconductor corresponds
 to a non-zero projection $l_z=\pm 1$ of the Cooper pairs angular momentum ${\bf {l}}$ along the $z$
 axis, and thus breaks time-reversal symmetry. The spatially homogeneous triplet order parameter
  $\hat\Delta=\Delta_0({\bf d}\cdot{\bf \hat\sigma})i\hat\sigma_y$ is described
 by the vector\cite{Pwave} ${\bf d}({\bf p})=(0,0,p_x+i\chi p_y)$, which depends on the
 direction of electron momentum ${\bf p}$. Here $\Delta_0$ is the bulk value of the order parameter,
 ${\bf \hat\sigma}=(\hat\sigma_x, \hat\sigma_y, \hat\sigma_z)$
  is the vector of Pauli matrices of conventional spin operators, and $\chi=\pm 1$
 corresponds to the two possible values of chirality. Also, chiral
 superconducting states can be associated with an admixture of two
 order parameters corresponding to different irreducible representations
 of a crystal point group. Naturally different order parameter
 components can coexist in the vicinity of interfaces
 between superconductors and surfaces due to broken symmetry of the crystal
 group \cite{Buchholtz1,Matsumoto1,Sigrist,Sigrist_PTP}. Among
 other possibilities of subdominant order parameter symmetry \cite{Buchholtz1} there
 are states which break time reversal symmetry
 \cite{Matsumoto1,Sigrist,Sigrist_PTP}. The coexistence of order
 parameters shows up in the local density of states\cite{Buchholtz2,Matsumoto2,Fogelstrom,Tanuma} as well as in
 the generation of spontaneous currents flowing along the surfaces
 in time reversal symmetry breaking cases\cite{Matsumoto2,Fogelstrom}.
 \par
 Time-reversal symmetry breaking order parameters have been proposed to exist near
surfaces \cite{High-Tc} and within vortex cores\cite{FT} in
high-$T_c$ superconductors. This proposal stems from the
observation of a split zero-bias conductance peak in the
\textit{absence} of any applied magnetic field. In this case, it
has been suggested that the
 relevant order parameter is either $d+\i s$- or $d+\i d'$- wave. The gap may then be written as
$\Delta = \Delta_0g(\theta_p) + \i\Delta_s$ or $\Delta =
\Delta_{0}g(\theta_p) + \i\Delta_{d}g_1(\theta_p)$,
 respectively, where $\Delta_0$ is an amplitude of the main
 component and the admixture of another pairing symmetry is
denoted by the amplitudes $\Delta_s$ and $\Delta_d$.  Here,
 $\theta_p$ is a polar angle in momentum space
${\bf p}=p(\cos\theta_p,\sin\theta_p)$, $g(\theta_p) =
\cos(2\theta_p+\alpha)$ and $g_1(\theta_p) =
\sin(2\theta_p+\alpha)$ where $\alpha/2$ is an angle measuring the
misorientation of crystalline symmetry axes and coordinate axes.
One obtains $d_{x^2-y^2}$-wave symmetry of the main order
parameter component for $\alpha=0$ and $d_{xy}$-wave pairing for
$\alpha=\pi/2$. While the experimental data so far clearly
indicates an order parameter which breaks time-reversal symmetry,
the question of whether the symmetry is $d+\i s$- or $d+\i
d'$-wave remains unsolved. Clearly, experimental signatures that
 may distinguish
these two types of pairings would be highly desirable.

One of the important features of unconventional superconductors is the possibility for the existence of
surface Andreev bound states \cite{Hu,Tanaka,yang_prb_94}. They occur in the vicinity of the scattering interface
 between a superconductor and an insulator, if the incident and
 reflected quasiparticles (QP) with different momentum directions see different phases of the order parameter.
 The consequence of the Andreev bound states formation is an increase
 of the local density of states at the surface resulting in zero-bias conductance peak anomaly
 observed \cite{wei_prl_98} in tunneling spectroscopy of high-$T_c$ cuprates with $d$-wave symmetry of
 superconducting pairing as well as in  \cite{mao_prl_01} the $p$-wave triplet superconductor
 Sr$_2$RuO$_4$. Also, the Andreev bound states determine the anomalous
 low--temperature behaviour of the London penetration length\cite{Barash_Penetration}
 and the Josephson critical current in $d$-wave \cite{Barash_Josephson} and chiral
 superconductors \cite{Barash_Chiral}.
\par
  Under the influence of an applied magnetic
 field, screening currents and vortices may be generated in a superconductor. As a result,
 the spectrum of surface states acquires a Doppler shift, leading to a splitting of the
 zero-bias conductance peak\cite{High-Tc}. Abrikosov vortices located near a superconducting surface
 generate an essentially inhomogeneous superfluid velocity
 field, which leads to a non-trivial electronic structure of the surface bound states
  \cite{graser_prl_04,graser_prb_05,yokoyama_prl_08}.
 Also, it was recently \cite{yokoyama_prl_08} proposed that
 the same Doppler shift effect should lead to a chirality selective
 influence of the magnetic field on the surface states in a $p$-wave chiral superconductor
  with broken time-reversal invariance. The quasiparticle density of states (DOS)
  near a flat surface was shown
 to depend on the orientation of magnetic field with respect to the chirality as well as on the
 vorticity in case when the Abrikosov vortex is pinned near the surface of
 superconductor. Additionally, in superconductors featuring gap
 nodes, such as in the case in pure $d_{x^2-y^2}$-symmetric superconducting cuprates,
 a vanishing pair potential in nodal directions results in important
 ramifications for the physics of the system \cite{FT,Kopnin-Volovik-1996,Maki,FTVM,Melnikov_Dwave}.

   To understand the effect of an externally applied magnetic field on the surface DOS,
   let us consider a spectrum of Andreev bound states near a flat surface of a
   time-reversal symmetry breaking superconductor occupying the half-space $x>0$.
    Below, we focus on the $p+\i p$-wave,  $d_{xy}+\i s$- and $d_{xy}+\i d_{x^2-y^2}$ -wave cases for concreteness.
    We consider a model situation assuming spatially homogeneous gap function, having
    the following form in momentum space:
 \begin{equation}\label{GapP}
 \Delta = \Delta_0e^{i\chi\theta_p}
 \end{equation}
 for $p+ip$-wave,
 \begin{equation}\label{GapD+iS}
 \Delta = \Delta_0\sin(2\theta_p) + \i\Delta_s
 \end{equation}
for $d_{xy}+is$- wave and
 \begin{equation}\label{GapD+iD}
 \Delta = \Delta_0\sin(2\theta_p) + \i\Delta_d\cos(2\theta_p)
 \end{equation}
for $d_{xy}+id_{x^2-y^2}$-wave superconductors.

    Assuming that the QP are specularly reflected at
 the surface of the superconductor within a Doppler shift approach\cite{Tinkham},
 the spectrum of the surface states can be expressed as follows\cite{Barash_Chiral, Sigr}:
$\varepsilon_a=\varepsilon_{a0}+\epsilon_d$, where
$\varepsilon_{a0}$ is a position of energy level in zero magnetic
field and $\epsilon_d=\hbar {\bf{k}}_F{\bf{v}}_s$ is the Doppler
shift energy
 which is determined by a local field of superfluid velocity.
The superfluid velocity ${\bf v}_s$ near a surface has only a
tangential
 component, directed along the $y$ axis ${\bf{v}}_s=(0,v_{sy},0)$,
 and can be related to the density of supercurrent flowing along the surface
${\bf j_s}=en{\bf v_s}$, where $e$ is electron charge and $n$ is a
concentration of Cooper pairs. The magnetic field is screened in a
superconductor at the London length $\lambda$ as follows
$B=He^{-x/\lambda}$, where $H$ is a value of magnetic field outside the superconductor.
Therefore, the superfluid velocity is
$v_{sy}=-(2e/mc)\lambda H$.
\par

 If the magnetic field is absent, the spectrum of surface states
 is given by \cite{Hu,Volovik}
 \begin{equation}\label{ABS-P}
 \varepsilon_{0a}=\chi\Delta_0 k_y/k_F
 \end{equation}
for a chiral $p$-wave,
 \begin{equation}\label{ABS-D+iS}
 \varepsilon_{a0}=\Delta_s \text{sgn}(k_y)
 \end{equation}
for a $d_{xy}+is$-wave and
\begin{equation}\label{ABS-D+iD}
\varepsilon_{a0}=\Delta_{d}\text{sgn}(k_y)\cos(2\theta_p)
\end{equation}
for a $d_{xy}+id_{x^2-y^2}$- wave.
  Here $k_y$ is a projection of QP momentum
along the surface. The above spectra may be formally obtained by solving \cite{yang_prb_94}
\begin{align}
\frac{\Delta(\theta_p)}{\varepsilon_{a0}-\i\sqrt{|\Delta(\theta_p)|^2-\varepsilon_{a0}^2}}
= \frac{\Delta(\pi-\theta_p)}{\varepsilon_{a0}+\i\sqrt{|\Delta(\pi-\theta_p)|^2-\varepsilon_{a0}^2}}.
\end{align}
In the $d_{x^2-y^2}+is$ wave case, one finds that
$\varepsilon_{a0} = \sqrt{\Delta_0^2\cos^2(2\theta_p)
 + \Delta_s^2}$, from which one infers that there are no subgap surface states. This is qualitatively
  different from the $d_{xy}+is$ wave case. The interesting effects occur in the latter case,
   so we focus on the $d_{xy}+is (d)$ wave symmetry in the following, corresponding to $\alpha=\pi/2$.
\par
\begin{figure}[h!]
\centering \resizebox{0.5 \textwidth}{!}{
\includegraphics{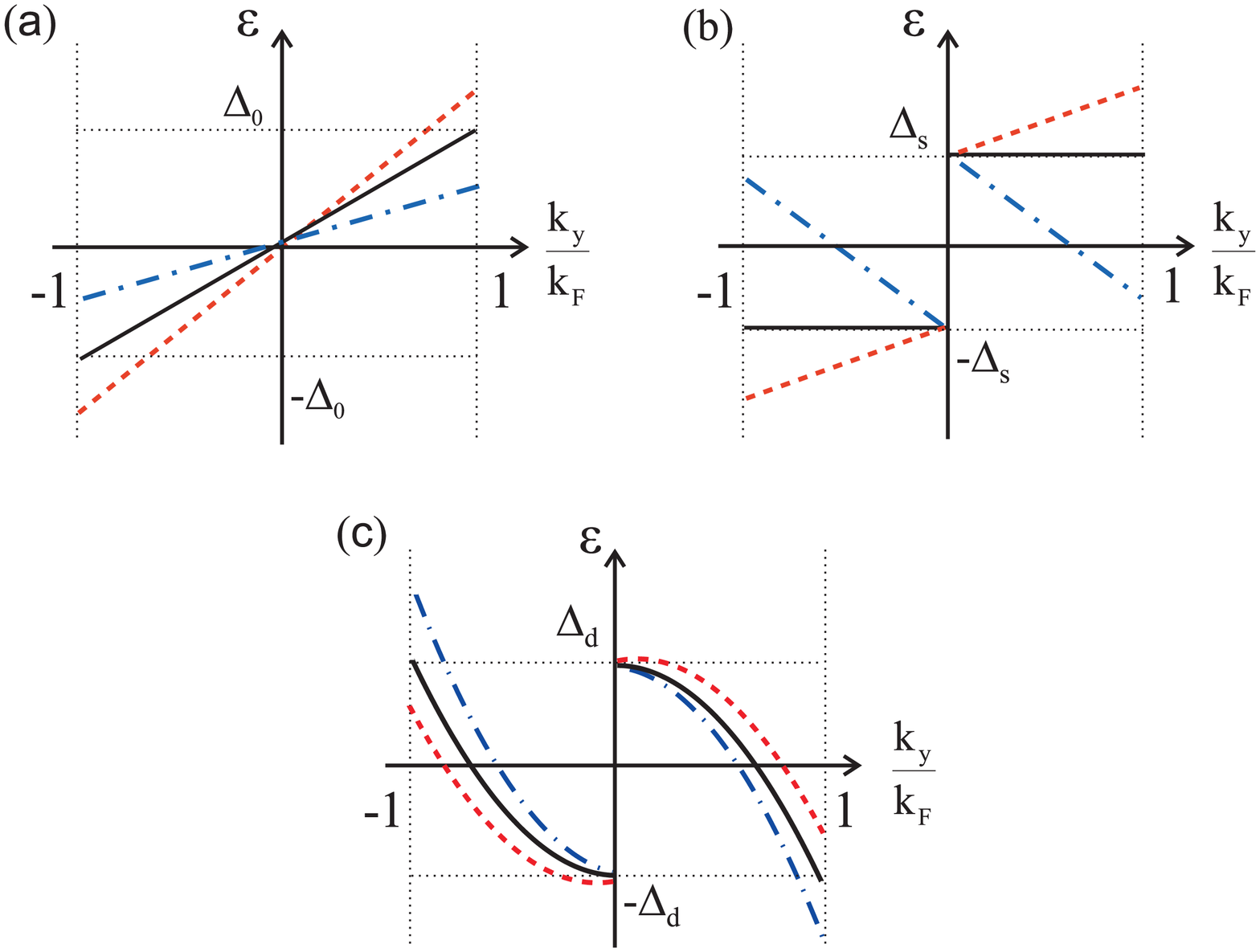}}
\caption{(Color online) Plot of the surface states spectrum for
(a) chiral $p+ip$-wave superconductor, (b) $d+is$- wave and (c)
$d+id$- wave superconductors. Spectrum in zero magnetic field is
shown by solid lines. Blue and red lines correspond to the
spectrum transformation due to magnetic field directed along and
opposite to $z$ axis correspondingly.} \label{SpectrumTr}
\end{figure}

The transformation of these spectra due to the Doppler shift
effect is shown in Fig.\ref{SpectrumTr}. To be definite we assume
that $\Delta_s>0$, $\Delta_{d}>0$ and $\chi=1$.  Considering the
DOS at Fermi level, $\nu=|\partial\varepsilon_a/\partial
k_y|^{-1}_{\varepsilon=0},$ in a chiral $p$-wave superconductor
one can see that its dependence on the magnetic field is
monotonic: it either increases or decreases for different field
directions (see Fig.\ref{SpectrumTr}a) as discussed in Ref.
\onlinecite{yokoyama_prl_08}.
\par
Another behaviour of the DOS occurs in the case of a $d+is$ wave
superconductor. From Fig. \ref{SpectrumTr}b it follows that for a
certain field direction there are no states at the Fermi level
$\varepsilon=0$ (red dashed lines in Fig.\ref{SpectrumTr}b). For
the opposite field direction (blue dash-dotted lines in
Fig.\ref{SpectrumTr}b), intersections of spectral branches with
the Fermi level appear when the superfluid velocity is large
enough $|v_{sy}|>\Delta_s/p_F$ so that the value of momentum
projection at the intersection point is smaller than the Fermi
momentum $|k_y^*|<k_F$. Thus, one can expect that the DOS at the
Fermi level should be zero when $H<H^*$, where $H^*$ is the
magnetic field value providing the condition
$|v_{sy}|=|\Delta_s|/p_F$ to be fulfilled.

On the contrary, in the $d+id$ wave case the DOS at the Fermi level is
non-zero even in the absence of a magnetic field. As can be seen
from Fig. \ref{SpectrumTr}c (black solid lines) the spectral
branches intersect the level $\varepsilon=0$ at $k^*_y=\pm
k_F/\sqrt{2}$. The transformation of the spectrum due to the magnetic
field of different directions is shown in Fig. \ref{SpectrumTr}c
by red dashed lines ($H>0$) and by blue dash-dotted lines ($H<0$).
Then, it can be easily seen that for $H>0$ the coordinates of the
intersection points $k^*_y$ shift towards $\pm k_F$ and for a
certain value of the magnetic field $H>H^*$ the DOS at the Fermi level
$\varepsilon=0$ disappears.

In the presence of an Abrikosov vortex near the surface of chiral
superconductor a non-trivial structure of the local density of
states distribution appears which depends on the vortex
orientation \cite{yokoyama_prl_08}.
  Along with the Doppler shift effect\cite{yokoyama_prl_08}, an important modification of the
 quasiparticle spectrum and the DOS can be obtained due to the overlapping of the surface
 states and the low-energy QP states localized within the vortex
 core, found in the pioneering work by Caroli- de
Gennes and Matricon (CdGM)\cite{CdGM}. It was shown that QP states
 with energy lower than the bulk superconducting gap value
 $\Delta$ are localized within the vortex core at the characteristic
  scale of the order of coherence length $\xi$  and have a
discrete spectrum $\varepsilon_v(\mu)$ as a function of the
quantized (half--integer) angular momentum $\mu$. At small
energies $|\varepsilon|\ll\Delta$ the spectrum for a vortex with
vorticity $M$ is given by
\begin{equation}\label{CdGMeq}
\varepsilon_v(\mu)\approx-M\mu\omega,
\end{equation}
where $k_F=p_F/\hbar$ and $\omega\sim \Delta_0 /(k_F\xi)$.
 For the most of superconducting materials, including Sr$_2$RuO$_4$, the interlevel spacing
 $\omega$ is much less than the superconducting gap $\Delta$ since
$(k_F\xi)\gg 1$. Therefore, the CdGM spectrum may be considered
continuous as a function of the impact parameter of the
quasiclassical trajectory $b=-\mu/k_F$ and the direction of QP
momentum $\theta_p$ as in the following form
\begin{equation}\label{CdGMQc}
\varepsilon_v(b,\theta_p)\approx M \Delta (b/\xi).
\end{equation}
 In the case of a chiral $p$-wave superconductor, the spectrum of
vortex core states differs from the CdGM result and is given by
Eq. (\ref{CdGMeq}) with integer $\mu$. For the $d+is$- and $d+id$-
wave superconductors the quasiclassical spectrum of vortex core
states is given by Eq.(\ref{CdGMQc}) with $\Delta (\theta_p)
=\sqrt{\Delta_0^2g^2(\theta_p)+\Delta_s^2}$ and $\Delta (\theta_p)
=\sqrt{\Delta_0^2g^2(\theta_p)+\Delta_d^2g_1^2(\theta_p)}$. The
discrete spectrum is obtained by applying Bohr-Sommerfeld
quantization rule to the canonical variables $\mu=-k_Fb$ and
$\theta_p$ \cite{Kopnin-Volovik-1996,Kopnin_BSrule}. It should be
noted that when the superconducting
 order parameter contains nodes, the quasiclassical expression (\ref{CdGMQc})
 is invalid near the nodal directions
 since energy states near the vortex core are
 not truly localized, but rather "leak" out through the gap nodes \cite{FT,Kopnin-Volovik-1996,Maki,FTVM,Melnikov_Dwave}.
 This is not the case for us since we consider superconducting order parameters
 which are gapped over the entire Fermi surface.

\begin{figure}[h!]
\centering \resizebox{0.5\textwidth}{!}{
\includegraphics{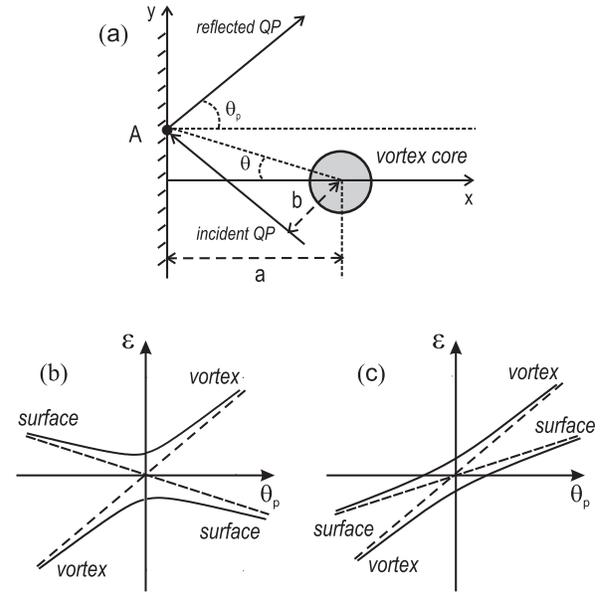}}
\caption{  Sketch of QP trajectories forming  surface and vortex
states (a) and qualitative plot of spectrum transformation due to
the interaction of surface and vortex states in the case of a
chiral $p$-wave superconductor; vorticity and chirality have (b)
equal and (c) opposite
  values.  } \label{SpectrumTrans}
\end{figure}

To study the interaction between vortex and surface states, let us
consider an example of vortex positioned near a flat surface of
chiral $p+ip$-wave superconductor. Comparing the energies of
surface $\varepsilon_a$ (\ref{ABS-P}) and vortex $\varepsilon_v$
(\ref{CdGMQc}) states one can see that for certain QP trajectories
the condition of resonance   $\varepsilon_a=\varepsilon_v$ is
realized. Thus the spectrum transformation in such almost
degenerate two-level system is given by a secular equation:
\begin{equation}\label{TwoLevel}
  (\varepsilon-\varepsilon_a)(\varepsilon-\varepsilon_v)=J^2.
\end{equation}
  Since we consider a low-energy spectrum $|\varepsilon|\ll
 \Delta_0$, the trajectories should pass close to the vortex center for
 the spectrum modification (\ref{TwoLevel}) to be effective. Then,
 the interaction of surface and vortex states
    is determined by the overlap integral $J\approx\Delta
    \exp(-\tilde{a}/\xi)$, where $\tilde{a}=a/\cos\theta_p$ and $a$
     is the distance from the vortex to the surface.
Taking a certain point at the surface (see point $A$ in
Fig.\ref{SpectrumTrans}a) of the superconductor one can obtain a
relation between the angles and impact parameters
 of trajectories passing trough this
point as follows $b=\tilde{a}\sin(\theta-\theta_p)$. Thus the
energy of vortex core states can be written
    as $\varepsilon_v= M(\tilde{a}/\xi) \Delta_0 \sin(\theta-\theta_p)$. Then, from
    Eq.(\ref{TwoLevel}) we obtain the spectrum transformation
    shown qualitatively in Fig.\ref{SpectrumTrans} for the particular case of $\theta=0$.
     It is easy
    to see that for equal values of vorticity and chirality
    (Fig.\ref{SpectrumTrans}b) there appears a minigap in
    quasiparticle spectrum near the Fermi level and therefore the zero-energy
    DOS is suppressed. On the other hand, in the case of opposite
    vorticity and chirality (Fig.\ref{SpectrumTrans}c) there is no minigap
    and the DOS is not suppressed.

In a $d+is$- and $d+id$- wave superconductor, the interaction
between vortex and surface states can also lead to noticeable
effects, which will be discussed later in the present paper.

Recently, it was pointed out that tunneling of quasiparticles into
vortex core states leads to a resonant enhancement of subgap
conductance of normal metal/superconductor (N/S) junction \cite{Silaev}. In the case of chiral
superconductors, such a tunneling effect can lead to either
stimulation or suppression of conductance, depending on the
direction of vorticity. We will show that if vortices are located
far from the N/S interface, the conductance follows the behaviour
expected from the Doppler shift approach. On the other hand, when the
distance from the vortex to the interface becomes comparable with coherence
length $\xi$ the tunneling into vortex core states comes into play,
leading to the peculiar nonmonotonic conductance dependence on the
vortex coordinate with respect to the superconducting surface.
\par
This paper is organized as follows. In Sec. \ref{sec:theory}, we
give an overview of the theoretical framework which is employed in
this work, namely a Bogolioubov approach and a quasiclassical
Eilenberger approach. In Sec. \ref{sec:results}, we present our
main results for the influence of magnetic field on bound surface
states spectra and local density of states near the surface. We
discuss the transformation of surface states in the Miessner state
of superconductor as well as the effects of interplay between
surface and vortex core states. We give our conclusions in Sec.
\ref{sec:summary}.

\section{Theoretical approach}\label{sec:theory}

Our further considerations are based on the Bogoliubov--de Gennes
(BdG) equations for particle-- ($u$) and hole--like ($v$) parts of
the wave function, which have the following form:
\begin{eqnarray}
\label{BdG}
  \nonumber
  -\frac{1}{2m}\left(\hat {\bf p}-\frac{e}{c}{\bf A}\right)^2u+\hat\Delta v
  &=&(\varepsilon+\varepsilon_F) u\,,\\
  \frac{1}{2m}\left(\hat {\bf p}+\frac{e}{c}{\bf A}\right)^2v
 +\hat\Delta^\dag u&
  =&(\varepsilon-\varepsilon_F) v\,.
\end{eqnarray}
Here $\hat\Delta$ is a gap operator, ${\bf A}$ is a vector
potential, $\hat {\bf
p}=-i(\partial/\partial_x,\partial/\partial_y)$, and ${\bf
r}=(x,y)$ is a radius vector in the plane perpendicular to the
anisotropy plane. Hereafter we assume the Fermi surface to be
cylindrical
 along the $z$-- axis and consider a motion of QPs only in $xy$ plane.

In case of unconventional superconductors, the gap potential
$\hat\Delta$ is a non-local operator, so the BdG system
effectively becomes a very complicated integro-differential
equation. Another complexity arises from the broken spatial
invariance of the superconducting gap in presence of vortices near
the N/S interface. A simplification can be obtained if one
considers a quasiclassical approximation, assuming that the
wavelength of quasiparticles is much smaller than the
superconducting coherence length (see e.g.
Ref.\onlinecite{Bardeen}). Within such an approximation, QP move
along linear trajectories, i.e. straight lines along the direction
of QP momentum
${\bf{n}}={\bf{k_F}}k_F^{-1}=(\cos\theta_p,\sin\theta_p).$
Generally, the quasiclassical form of the wave function can be
constructed as follows: $(u,v)=e^{i{\bf{k_F}}\cdot{\bf{r}}}(U,V)$,
where $(U({\bf{r}}),V({\bf{r}}))$ is a slowly varying envelope
function. Then the system (\ref{BdG}) reduces to a system of
first-order differential equations along the linear trajectories
defined by the direction of the QP momentum
${\bf{n}}={\bf{k_F}}k_F^{-1}=(\cos\theta_p,\sin\theta_p).$
Introducing the coordinate along trajectory
$x'=({\bf{n}}\cdot{\bf{r}})=r\cos(\theta_p-\theta)$ we arrive at
the following form of the quasiclassical equations:
\begin{align}\label{quasiclass}
  \left(-i\hbar v_F\partial_{x'}+\mathbf{v}_F\cdot\frac{e}{c}\mathbf{A}\right)U+\Delta
  V=\varepsilon U \notag \\
\left(i\hbar
v_F\partial_{x'}+\mathbf{v}_F\cdot\frac{e}{c}\mathbf{A}\right)V+
\Delta^\dag U=\varepsilon V,
\end{align}
where the Fermi velocity is $\mathbf{v}_F={\bf{n}}\hbar k_F/m$.
The pairing potential in Eq. (\ref{quasiclass}) may generally be
written as
\begin{align}
\Delta(\mathbf{r},\theta_p) = \Delta(\theta_p)\Psi(\mathbf{r}),
\end{align}
where $\Delta(\theta_p)$ describes the orbital symmetry of the
superconducting order parameter in momentum space, while
$\Psi(\mathbf{r})$ describes its spatial dependence both
magnitude- and phase-wise.

The local DOS (LDOS) can be expressed through the eigenfunctions of the
BdG equation (\ref{BdG}) in the following form \cite{Mahan}:
\begin{equation}\label{DOS}
  N(\varepsilon,{\bf{r}})=\sum_n
  |u_n({\bf{r}})|^2\delta(\varepsilon-\varepsilon_n),
\end{equation}
 where $u_n({\bf{r}})$ is electron component of
 quasiparticle eigen function corresponding to an energy
 level $\varepsilon_n$. The eigenfunction has to be normalized:
 $\iint_{-\infty}^{\infty} |u_n ({\bf r})|^2+|v_n ({\bf r})|^2 d^2r=1$.
\par
We will also later employ a quasiclassical Eilenberger approach to study the spatially resolved DOS.
Let us here sketch the framework of the treatment which makes use of the Eilenberger equation, following
 the notation of Refs. \cite{schopohl_prb_95, dahm_prb_02}.
It is now convenient to solve the Eilenberger equation along
trajectories along the Fermi momentum, and introducing a
Ricatti-parametrization for the Green's function
\cite{schopohl_prb_95}. In this way, one obtains
\cite{graser_prl_04}
\begin{align}\label{eq:eilenberger}
\hbar v_F\partial_{x'}a(x') &+ [2\tilde{\omega}_n + \Delta^\dag
a(x')]a(x') - \Delta = 0,\notag\\
  \hbar v_F\partial_{x'}b(x') &-
[2\tilde{\omega}_n + \Delta b(x')]b(x') + \Delta^\dag = 0,
\end{align}
where $\i\tilde{\omega}_n = \i\omega_n +
m\mathbf{v}_F\cdot\mathbf{v_s}$ is a Doppler-shifted Matsubara
frequency and
 $$\mathbf{v_s}=\frac{1}{2m}\left(\hbar\nabla\Phi-\frac{2e}{c}\mathbf{A}\right)$$ is a gauge-invariant
  superfluid velocity where $\Phi (\mathbf{r})$ is a gap function
  phase: $\Psi (\mathbf{r})= |\Psi|e^{i\Phi}$.
  The LDOS may be expressed through the
scalar coherence functions $a$ and $b$ as follows
\cite{graser_prl_04}
\begin{align}\label{eq:LDOS}
N(\varepsilon) = \int^{2\pi}_0 \frac{\text{d}\theta}{2\pi} \text{Re}\Big\{\frac{1-ab}{1+ab}\Big\}_{\i\omega_n
\to \varepsilon +\i\delta},
\end{align}
where $\varepsilon$ is the quasiparticle energy measured from Fermi level and $\delta$ is a scattering parameter
which accounts for inelastic scattering.
\par
To investigate the transport properties of N/S junction, we employ
an approach similar to what was used in work by Bardeen,
Tinkham and Klapwijk \cite{BTK}.
 The expression for
the dimensionless zero- bias conductance of the N/S junction
measured in terms of the conductance quantum $e^2/(\pi\hbar)$ can
be written as follows:
\begin{equation}\label{cond1}
  G= \frac{G_{sh}}{2}\int_{-\pi/2}^{\pi/2}\left[1-R_n(\theta_0)+R_a(\theta_0)\right]\cos\theta_0\,d\theta_0,
\end{equation}
where $R_n(\theta_0)$ and $R_a(\theta_0)$ are the probabilities of
normal and Andreev reflection respectively, $\theta_0$ is the
incident angle: ${\bf{k_F}}=k_F(\cos\theta_0,\sin\theta_0)$,
characterizing the propagation direction of quasiparticles, coming
from the normal metal region. The Sharvin conductance
$G_{sh}=k_FL_y/\pi$ equals the total number of propagating modes
determined by the channel width $L_y$.

The problem of quasiparticle scattering at the N/S interface is
formulated within the BdG theory
(\ref{BdG}).
An interfacial barrier separating the N and S regions
can be modeled by repulsive
  delta function potential $W(x)=W_0\delta(x)$, parameterized by a dimensionless barrier strength $Z=W_0/\hbar v_F$.
    The boundary conditions at the N/S interface then read:
 \begin{align}\label{bc}
  \left[f (0)\right]=0,\;\;\;
  \left[\partial_x f(0)\right]=(2k_FZ) f(0),
 \end{align}
  where $f=(u,v)$ and $[f(x)]=f(x+0)-f(x-0)$.

 Considering
 a zero-bias problem we will have to analyze only zero-energy
excitations with $\varepsilon=0$. For wave functions in S region
corresponding to subgap quasiparticles, the following
  representation can be used: $(U,V) =e^{\zeta}\left(e^{i(\eta+\Phi)/2},
  e^{-i(\eta+\Phi)/2}\right)$, where $\zeta =\zeta (s,b)$
  and $\eta =\eta (s,b)$ are real-valued functions. Then, the quasiclassical
  equation (\ref{quasiclass})
  can be written as follows:
  \begin{align}\label{eta}
\partial_{x'}\eta+2|\Delta|\cos\eta+2\epsilon_d=0,\\
\partial_{x'}\zeta+2|\Delta|\sin\eta=0\notag.
\end{align}
where $\epsilon_d({\bf{r}})=\hbar{\bf{k}_F}{\bf{v_s}}$ is a
 Doppler shift energy. For wave functions decaying at the different ends
 of trajectory $(U,V)(x'=\pm\infty)=0$ from Eqs.(\ref{eta}) we obtain:
\begin{equation}\label{tetaBC}
\eta(x'=\pm\infty)=\pm\pi/2.
\end{equation}

 The boundary conditions (\ref{bc}) model the specularly reflecting N/S interface,
  coupling the waves with wave vectors ${\bf{k}}_F=k_F(\cos\theta_0,\sin\theta_0),$
  and ${\bf{k^\prime}}_F=k_F[\cos(\pi-\theta_0),\sin(\pi-\theta_0)]$. Therefore
 if the incident electron wave is $u_i=e^{i{\bf{k_F}}{\bf{r}}},$
 then the reflected electron $u_r$ and hole $v_r$ waves will have the form
 $$u_r=U_re^{i{\bf{k^\prime_F}}{\bf{r}}},\;\;v_r=V_re^{i{\bf{k_F}}{\bf{r}}},
$$
 where $U_r$ and $V_r$ are the envelope
 functions. Thus, each point $(0,y)$ at the N/S interface
 lies on the intersection of two quasiclassical trajectories, characterized by the
 angles $\theta_p=\theta_0$ and $\theta_p=\pi-\theta_0$. Let us
 denote the distribution of phases $\eta(x')$ along these
 trajectories as $\eta_+(x')$ and $\eta_-(x')$ correspondingly.
 Using the boundary conditions we obtain the following expression
 for the conductance\cite{Silaev}:
\begin{equation}\label{cond0}
  G=\frac{N_0}{2}\int^{L_y/2}_{-L_y/2} \int_{-\pi/2}^{\pi/2}
 g(y,\theta_0)\cos\theta_0\,d\theta_0 dy,
\end{equation}
where $g(y,\theta_0)$ is given by
 \begin{equation}\label{cond2}
  g(y,\theta_0)=\frac{2}{(\tilde{Z}^4+\tilde{Z}^2)|1-e^{i\rho}|^2+1},
\end{equation}
with $\tilde{Z}=Z/\cos\theta_0$ and
$\rho(y,\theta_0)=\eta_--\eta_+$ is determined by a difference of
phases $\eta_-(x')$ and $\eta_+(x')$ at the intersection point
$(0,y)$.
 To evaluate the conductance, one needs to find the factor
 $e^{i\rho}$ in Eq.(\ref{cond2}) and then the reflection probabilities
 by solving numerically Eq.(\ref{eta})
 with the boundary conditions in Eq. (\ref{tetaBC}).

\section{Results}\label{sec:results}

 To illustrate the basic effect of how the interplay between
the Doppler-shift and the time-reversal symmetry breaking of the
superconducting order parameter is manifested, we consider a
situation where an external magnetic field is applied near the
surface of the superconductor along the $\hz$-axis, thus inducing
a vector potential $\mathbf{A}$ in the superconductor which drives
the shielding supercurrent. In order to proceed analytically, we
make the simplifying assumption that the superfluid velocity field
is nearly homogeneous and that the spatial variation of the
superconducting order parameter near the
  interface is small. Choosing a real gauge, we then find that the Ricatti-functions $a$ and $b$ in
   Eq. (\ref{eq:eilenberger}) may be written as: \cite{graser_prl_04, yokoyama_prl_08}
\begin{align}
a(\theta) &= s(\theta)\Delta(\theta),\; b(\theta) =
s(\theta)\Delta^*(\theta),\notag\\ s(\theta) &=
1/[\tilde{\omega}_n(\theta) + \sqrt{(\tilde{\omega}_n(\theta) )^2
+ |\Delta(\theta)|^2}],
\end{align}
where $\tilde{\omega}_n$ depends on $\theta$ through the
Doppler-shift. To evaluate the LDOS in Eq. (\ref{eq:LDOS}) at the
surface, we need to take into account proper boundary conditions
at $x=0$. Assuming an impenetrable surface with perfect
reflection, these boundary conditions read:
\begin{align}\label{eq:bc}
a_\text{surface}(\theta) = a(\pi-\theta),\;
b_\text{surface}(\theta) = b(\theta).
\end{align}
Inserting this into the expression for the LDOS, we obtain
\begin{align}\label{eq:n1}
N(\varepsilon) = 2\text{Re}\Big\{ \Big\langle
\frac{1}{1+a(\pi-\theta)b(\theta)} \Big\rangle
\Big\}_{\i\omega_n\to
 \varepsilon+\i\delta} - 1.
\end{align}
The $\langle\ldots\rangle$ denotes angular averaging, which we
restrict to angles $-\pi/2\leq\theta\leq\pi/2$
 due to the surface. It may be shown that for a chiral $p$-wave superconductor \cite{yokoyama_prl_08},
  the zero-energy DOS at the surface reads
 \begin{align}
 N(0) = 1 + \frac{\hbar k_Fv_{sy}}{\Delta_0} + \ldots,
 \end{align}
 while for pure $s$- or $d$-wave superconductors one finds
 \begin{align}
 N(0) = C_1 + C_2v_{sy}^2 + \ldots,
 \end{align}
 where $C_1$ and $C_2$ are arbitrary constants.
From numerical investigations of Eq. (\ref{eq:n1}) at
$\varepsilon=0$, we find that the zero-energy DOS may quite
generally be written as
 \begin{align}
 N(0) = C_1 + C_2v_{sy}  + \ldots,
 \end{align}
whenever the superconducting order parameters \textit{i)} break
time-reversal symmetry and \textit{ii)} support the presence of
subgap surface-bound states. This is the case both for the
$p_x$+$\i p_y$-wave pairing which is believed to be realized in
Sr$_2$RuO$_4$, as well as the $d$+$\i s$-wave and $d+\i d$-wave
pairings that are relevant for the cuprates. In particular,
tunneling spectroscopy measurements have
     indicated the presence of such a time-reversal symmetry breaking order parameter near
      surfaces by a split zero-bias conductance peak that was observed in the \textit{absence}
      of an external field in several experiments.
       \cite{High-Tc}.
\par\par

 In Ref. \onlinecite{dahm_prb_02}, it was pointed out that the
neglect of the gradient term in the Eilenberger equation is
expected to be a reasonable approximation as long as the
Doppler-shift energy $m\mathbf{v}_F\cdot\mathbf{v}_s$ is
  small compared to the local gap energy $\Delta(\theta)$. This approximation would then fail close to the vortex core
   or gap nodes of $\Delta(\theta)$. Nevertheless, in the model
   case of spatially homogeneous gap function and superfluid
   velocity field the gradient terms in Eilenberger equation can
   be neglected in the whole range of Doppler shift energies.
   However considering a model situation the above discussion nevertheless serves to illustrate our main
qualitative argument: namely, that chirality-sensitive effects
should
 be expected in superconductors with order parameters that \textit{i)} break time-reversal symmetry and \textit{ii)}
 support the presence of subgap surface-bound states.
 We now proceed to discuss the cases of $p_x+\i p_y$-wave and
 $d+\i s (d)$-wave pairing in more detail, since these are relevant to actual materials.

 \subsection{Surface states in  $p+ip$ and $d+is(d)$ superconductors
 under the influence of magnetic field.}

\begin{figure}[h!]
\centering \resizebox{0.48\textwidth}{!}{
\includegraphics{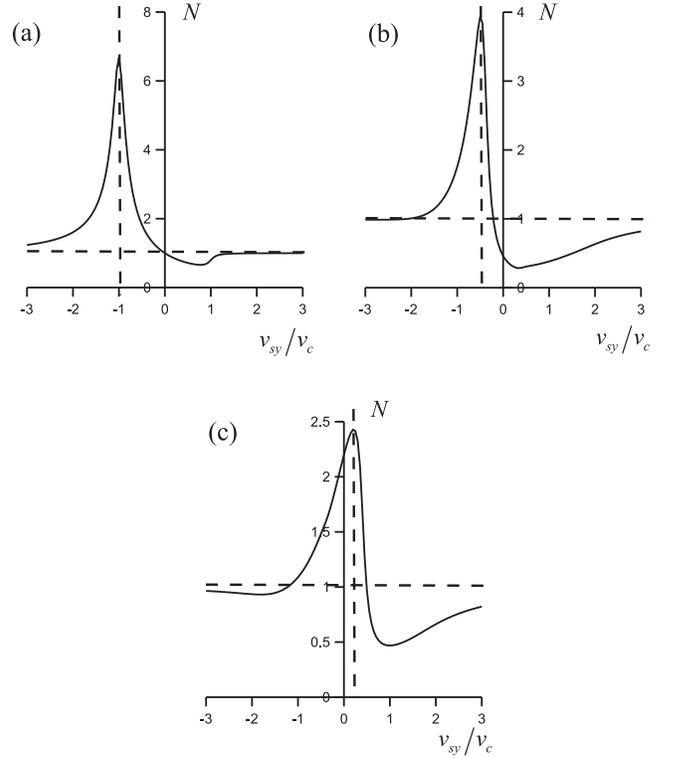}}
\caption{ Plot of the normalized zero-energy LDOS $N(0)$ for (a)
p-wave superconductor with $\chi=1$, (b) $d$+$\i s$-wave case with
$\Delta_s=0.4\Delta_0$ and (c) $d$+$\i d$-wave case with
$\Delta_{d}=0.4\Delta_{0}$. Dashed lines are guides for eyes: the
vertical ones denote positions of LDOS peaks and horizontal ones
correspond to the level of normal metal DOS $N_0$.}
\label{fig:DOSd+isp}
\end{figure}

 In Fig.\ref{fig:DOSd+isp}, we
 show numerical plots of the surface LDOS at the Fermi level given by Eq.(\ref{eq:n1}) for the chiral $p$-wave
 (Fig.\ref{fig:DOSd+isp}a), $d+is$-wave (Fig.\ref{fig:DOSd+isp}b) and
 $d+id$-wave (Fig.\ref{fig:DOSd+isp}c) cases in a wide domain of superfluid velocities.
  The structure of gap functions is chosen in the form of Eqs.
 (\ref{GapP})-(\ref{GapD+iD}) and the parameter characterizing inelastic scattering in Eq.(\ref{eq:LDOS}) is
 chosen as
 $\delta=0.1\Delta_0$.
We introduce the following notations for the different critical
velocities: $v_c=\Delta_0/(\hbar k_F)$, $v_{cs}=|\Delta_s|/(\hbar
k_F)$ and $v_{cd}=|\Delta_{d}|/(\hbar k_F)$.

 As seen, the surface LDOS has sharp peaks
 at a certain value of the superfluid velocity in all cases. We will show below
 that peaked structure of LDOS is provided by bound surface
 states.
Another contribution to the LDOS
 comes from the delocalized states corresponding to the continuous part of QP
 spectrum. A delocalized state with zero energy $\varepsilon=0$
 exists provided (i) $|v_{sy}|>v_c$  in case
 of chiral $p$-wave superconductor, (ii) $|v_{sy}|>v_{cs}$
 and (iii) $|v_{sy}|>v_{cd}$
 in
 case of $d+is$-wave and $d+id$-wave superconductors correspondingly.
 Condition (i) is unlikely
 to be realized because it means that the superfluid velocity is larger
 than the critical depairing value. Conditions (ii) and (iii) can be
 realized, because the values $v_{cs}$ and $v_{cd}$ can be well below the
 critical depairing velocity if the amplitude of additional order parameter components
  is small enough.

To analyze the contribution to LDOS provided by the bound surface
states we will consider the domain of low energies
$|\varepsilon|\ll \Delta_0$. By neglecting small deviations of
the electron and hole momentum, the normalized wave function of QP
localized near the boundary can be written as
 $$
 \left(u\atop v\right)=\left(1\atop i\right)\sqrt{\frac{2}{\tilde{\xi}|\cos\theta_p|}}
  e^{ik_yy}\sin(k_xx)e^{-x/(\tilde{\xi}\cos\theta_p)},
 $$
 where $(k_x,k_y)=k_F(\cos\theta_p,\sin\theta_p)$. This wave function decay in the
 superconducting side $x>0$ at a characteristic localization scale $\tilde{\xi}$ is given by
$\tilde{\xi}=\hbar v_F/\Delta_0$ for chiral
 $p$-wave and $\tilde{\xi}=\hbar
v_F/|\Delta_0\sin(2\theta_p)|$ for $d+is$- and $d+id$- wave
superconductor correspondingly with gap functions given by
Eqs.(\ref{GapD+iS},\ref{GapD+iD}).
 The spectrum of the Andreev bound states, shifted
 by the superfluid velocity, is given by
 \begin{equation}\label{DopplerABS-P}
  \varepsilon_a=\chi \Delta_0 k_y/k_F+\hbar v_{sy}k_y
 \end{equation}
 for the $p$-wave case,
 \begin{equation}\label{DopplerABS-D+iS}
 \varepsilon_a=\Delta_s \text{sgn}(k_y)+\hbar v_{sy}k_y
 \end{equation}
 for $d+is$-wave case and
  \begin{equation}\label{DopplerABS-D+iD}
 \varepsilon_a=\Delta_{d}\text{sgn}(k_y)\cos(2\theta_p)+\hbar v_{sy}k_y
 \end{equation}
 for $d+id$-wave case. Consequently, the contribution from Andreev bound states to the zero-energy LDOS at the
 surface of a chiral $p$-wave superconductor is given by
 $$
 N_a=N_0 \frac{1 }{|v_{sy}/v_c+\chi|},
 $$
 where $N_0 = m/(2\pi\hbar^2)$ is the normal metal LDOS per one spin direction.
 For a chiral $d+is$ superconductor, the behaviour of the LDOS is more
 complicated. Assuming that $\Delta_s>0$, we obtain that the LDOS is zero for $v_{sy}>-\Delta_s/{\hbar k_F}$.
 Otherwise, it is given by
 $$
 N_a=4 N_0 \frac{v_{cs}v_{c}}{v_{sy}^2}.
 $$
On the contrary, for the $d+id$ case the LDOS is zero if
$v_{sy}>\Delta_{d}/(\hbar k_F)$ (for $\Delta_d>0$) and otherwise
it is given by
 $$
 N_a= N_0
 \frac{\Delta_0}{|\Delta_d|}\left(1+\frac{v_{cd}}{\sqrt{v_{sy}^2+8v_{cd}^2}}\right).
 $$

 It can be seen that these
contributions to LDOS have peaks at $v_{sy}=v_{c}$ for $p$-wave
case. For $d+is$-wave and $d+id$-wave superconductors the peaks
are positioned at $v_{sy}=-sgn(\Delta_s)v_{cs}$ and
$v_{sy}=sgn(\Delta_d)v_{cd}$ correspondingly. Even though the position of the peaks are different, the dependencies of the surface LDOS on the superfluid
velocity (and consequently on magnetic field) are very similar for
$d+is$ and $d+id$-wave superconductors. Therefore,
it might be difficult to distinguish which case is realized experimentally.

On the other hand the considered model with a spatially
    homogeneous gap function $\Psi({\bf r})=1$
 is adequate only when the applied magnetic field is not too large.
 When the magnetic field is large enough, it breaks
the Meissner state and generates vortices near the surface of
superconductor. Therefore, we investigate the influence of
vortices on the LDOS distribution near a superconducting surface as
well as on the conductance of normal metal/superconducting
junctions. We will show that vortices have different effect on the
conductance in $d+is$ and $d+id$ cases.

\subsection{Interplay of vortex and surface states in chiral superconductors.}\label{sec:vortex}

  A chirality sensitive
 LDOS transformation due to vortices situated near the surface
 of a chiral $p$-wave superconductor was considered in
 Ref.\onlinecite{yokoyama_prl_08}. It was shown that depending on the chirality
 and vorticity
 value, the surface LDOS near is either enhanced or suppressed upon decreasing the
 distance from the vortex to the surface. In case of $d+is(d)$ superconductors the
 transformation of LDOS profile is also sensitive to the value of vorticity.
  Similar behaviour is expected
 for a conductance of normal metal / chiral superconductor
 junction in the presence of vortices.
 \par
  To investigate the influence of a single vortex
 on the LDOS profile and conductance, we assume that at
  $x>0$ (superconducting region) the coordinate dependence
  of the order parameter may be written as follows:
\begin{equation}\label{OP}
  \Psi({\bf{r}})=e^{i\Phi}.
\end{equation}
Here, we consider a model situation where the magnitude of the order
parameter is constant.
 The phase  distribution $\Phi({\bf{r}})$  consists of a singular part
$\Phi_v({\bf{r}})=arg({\bf{r}}-{\bf{r}}_v)$ and a regular part
$\Phi_r({\bf{r}}),$ determined by the particular metastable vortex
lattice configuration realizing near the boundary.
 We assume that the
 regular part of the phase distribution is $\Phi_r({\bf{r}})=-arg({\bf{r}}-{\bf{r_{av}}})$
 corresponding to the image vortex situated at the point ${\bf{r_{av}}}=(-2a,0,0)$
 behind the N/S interface.

 \subsubsection {$p+ip$ wave}

\begin{figure}[h!]
\centering \resizebox{0.5\textwidth}{!}{
\includegraphics{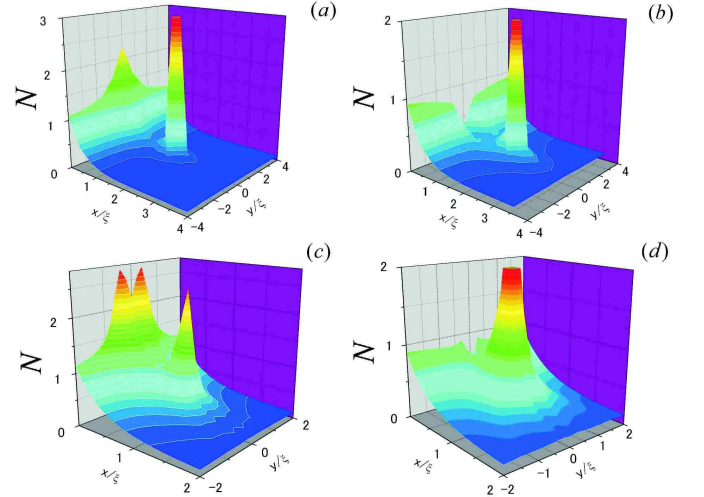}}
\caption{(Color online) Plot of the normalized zero-energy LDOS
$N(0)$ in the presence of a vortex near the surface of a chiral $p$-wave
superconductor. (a) and (c) correspond to equal vorticity and
chirality, (b) and (d) correspond to opposite vorticity and
chirality. The distance from vortex to the surface is $a=2\xi$ for
(a) and (b) and $a=\xi$ for (c) and (d).} \label{fig:LDOSp}
\end{figure}

In Fig.\ref{fig:LDOSp} we show the LDOS profile near the surface of a
chiral $p$-wave superconductor in the presence of a single vortex,
positioned at some distance $a$ from the surface. When the vortex is
positioned far from the surface $a\geq 2\xi$ the LDOS profile
follows the behaviour, expected from the picture of local Doppler
shift\cite{yokoyama_prl_08}. Depending on the relative value of
vorticity and chirality, the surface LDOS is either increased
(Fig.\ref{fig:LDOSp}a) or decreased (Fig.\ref{fig:LDOSp}b). An
analytical estimate with the help of spectrum Eq.
(\ref{LocalDoppler}) yields a following estimation of the
amplitude of LDOS peak in Fig.\ref{fig:LDOSp}a: $\Delta
N/N_0=(1+M\chi a)^{-1}$. At smaller distances $a\leq 2\xi$, the
behaviour of LDOS changes drastically. In the case of opposite
vorticity and chirality, the surface LDOS grows at $a\leq 2\xi$,
obviously due to the overlapping with the peak of vortex core
states. In case of equal vorticity and chirality the same
overlapping occurs, but on the contrary it leads to reduction of
DOS, as it was discussed in Sec. \ref{sec:introduction}. The peak
of the LDOS at the surface discussed in
Ref.\onlinecite{yokoyama_prl_08} transforms into a dip-and-peak
structure as the vortex comes close to the surface.

\begin{figure}[h!]
\centering \resizebox{0.45\textwidth}{!}{
\includegraphics{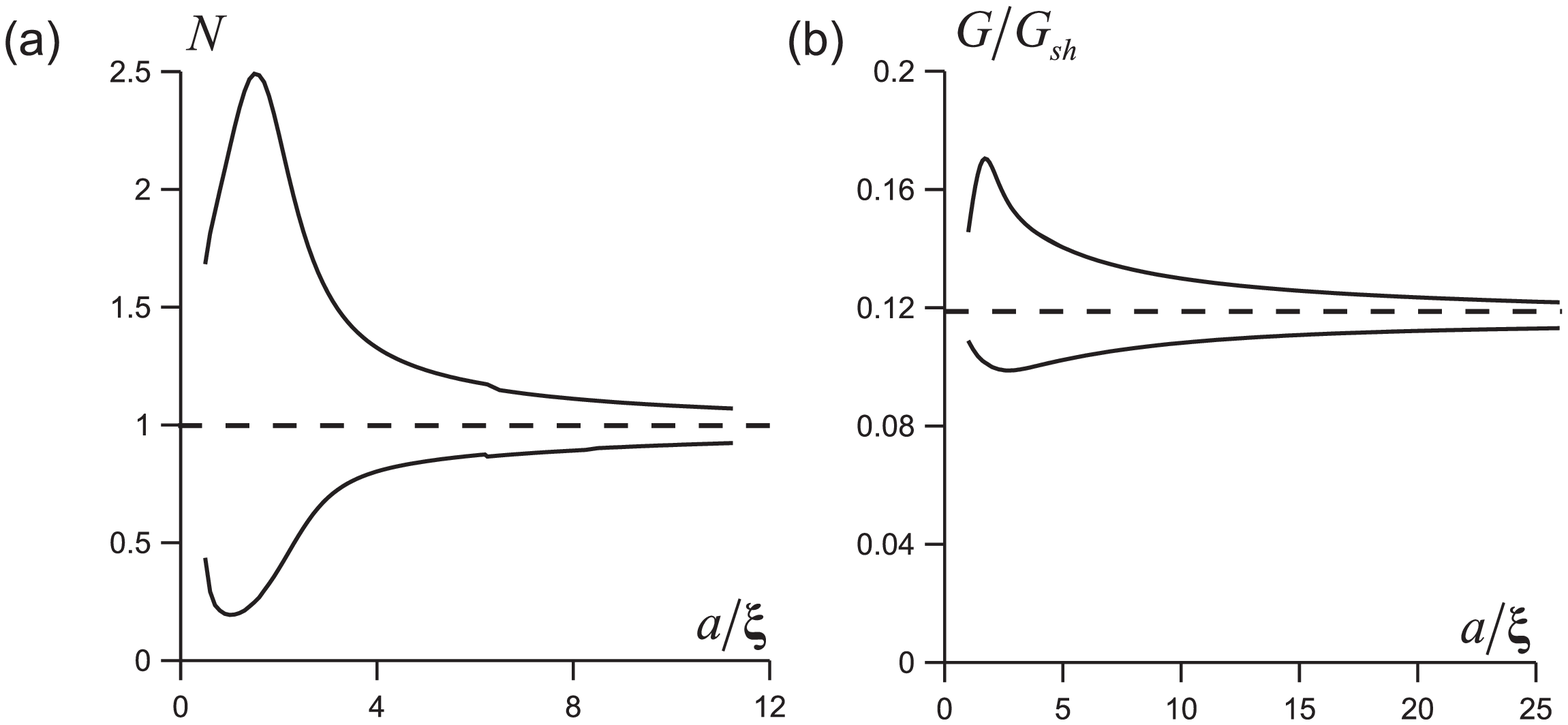}}
\caption{ (a) Plot of the normalized zero-energy LDOS at the point
on the surface which is closest to the vortex core. Different
curves correspond to different vorticities. (b) Plot of the
vortex-induced conductance
 in chiral $p+ip$-wave superconductor for equal and opposite
  values of vorticity
and chirality. The strength of interface barrier is $Z=5$. Large
distances asymptotics for $N$ and $G$ are shown by dash lines. }
\label{fig:Cond_Pwave}
\end{figure}

This is illustrated in Fig.\ref{fig:Cond_Pwave}(a), where we plot
the LDOS at the surface point $(0,0)$, which is the nearest point
to the vortex in Fig.\ref{fig:LDOSp}. At large distances $a\gg\xi$
the LDOS is a monotonic function of $a$, either increasing or
decreasing depending on the relation between vorticity and
chirality. At smaller distances $a\leq 2\xi$, the extremum of LDOS
appears. In the case of opposite vorticity and chirality [lower
curve in Fig.\ref{fig:Cond_Pwave}(a)], the surface LDOS grows at
$a\leq 2\xi$, due to the overlapping with the peak of vortex core
states. In case of equal vorticity and chirality [upper curve in
Fig.\ref{fig:Cond_Pwave}(a)] the same overlapping leads to
reduction of LDOS.

To investigate the influence of vortices on the transport
properties of normal metal/ chiral $p$- wave superconductor
junction we solve the generic problem of the influence of a single
vortex near the N/S surface on the zero-bias conductance of the
junction. A numerical plot of the conductance $G$ as a function of
a distance of vortex to the junction interface is shown by the solid
lines in Fig.\ref{fig:Cond_Pwave} (b) for equal (upper curve)  and
opposite (lower curve) values of chirality and vorticity. The
conductance is normalized to the value of Sharvin conductance
$G_{sh}=k_FL_y/\pi$.

   At large distances $a\gg \xi$ an analytical estimation of conductance
   can be obtained by using a local Doppler shift approximation on the quasiparticle spectrum.
   Indeed, the modification of the surface states energy due to a
   supercurrent flowing along the boundary of superconductor can
   be written as
\begin{equation}\label{LocalDoppler}
  \varepsilon_a\approx\left(\chi\Delta_0+\hbar v_{sy} k_F\right)k_y/k_F,
\end{equation}
   where $k_y$ is a quasiparticle momentum along the surface, $\chi=\pm 1$
  is a chirality value and $v_{sy}=(M\hbar/m) a/(y^2+a^2)$ is a projection on the surface plane of superfluid
  velocity generated by the vortex and image
  antivortex, $M$ is vorticity
 value and $m$ is the electron mass. It follows from Eq.(\ref{LocalDoppler}) that the Doppler
 shift effect leads to a change in the slope of anomalous branch.
 It is easy to obtain that in this case the function
 $g(y,\theta_0)$ in expression Eq. (\ref{cond2}) takes the following form:
 \begin{equation}\label{ConductanceApproximated}
  g(y,\theta_0)=\frac{2}{4(\tilde{Z}^4+\tilde{Z}^2)(
  \varepsilon_a/\Delta_0)^2+1}.
 \end{equation}
  The straightforward integration in Eq.(\ref{cond0}) yields $G=G_0+\delta
  G$, where $G_0=G_{sh}(\pi/Z^{2})$ is the conductance without
  vortex and
 \begin{equation}\label{VortexConductance}
 \delta G/G_{sh}=\pm \frac{2\pi\xi}{Z^{2}L_y}\arctan(L_y/2a)
 \end{equation}
 is a vortex-induced conductance shift,
 where the upper (lower) sign corresponds to equal (opposite)
 vorticity and chirality.

At distances smaller than $2\xi$, an extremum of the conductance
appears. Upon placing the vortex closer to the surface, an
opposite effect occurs: one obtains a conductance suppression
instead of enhancement and vice versa. The origin of the
conductance extremum is a tunneling of quasiparticles into the
vortex core states, or in other words, the overlapping of vortex
and surface bound states. Comparing Figs.\ref{fig:Cond_Pwave}(a)
and \ref{fig:Cond_Pwave}(b) one can see that the conductance in
general follows the behaviour of the surface DOS.

\subsubsection{$d+is$ and $d+id$ wave.}

In chiral $d+is$ and $d+id$ superconductors the LDOS
transformation appears to also be vorticity sensitive.
\begin{figure}[h!]
\centering \resizebox{0.5\textwidth}{!}{
\includegraphics{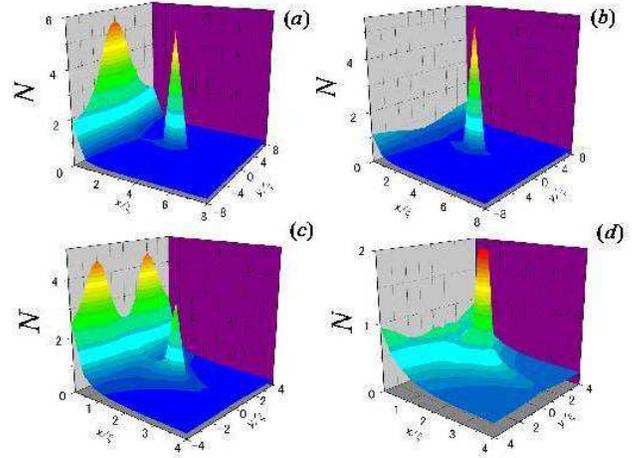}}
\caption{(Color online) Plot of the normalized zero-energy LDOS
profile $N(0)$ in the presence of vortex near the surface of
chiral $d+is $-wave superconductor with $\Delta_s=0.2\Delta_0$.
  (a),(c) and (b),(d) correspond to different vortex orientations with respect to $z$ axis.
   The distance from vortex to the surface is $a=4\xi$ for
(a) and (b) and $a=2\xi$ for (c) and (d). } \label{fig:LDOSd+is}
\end{figure}
\begin{figure}[h!]
\centering \resizebox{0.5\textwidth}{!}{
\includegraphics{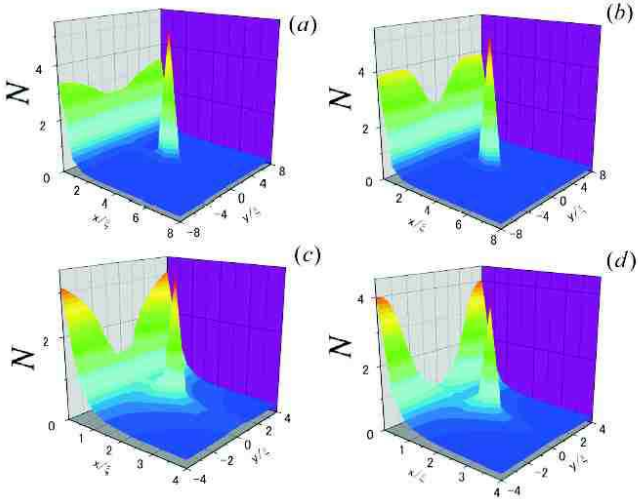}}
\caption{(Color online) Plot of the normalized zero-energy LDOS
profile $N(0)$ in the presence of vortex near the surface of
chiral $d+id $-wave superconductor with $\Delta_d=0.2\Delta_0$
(a),(c) and (b),(d) correspond to different vortex orientations
with respect to $z$ axis.
   The distance from vortex to the surface is $a=4\xi$ for
(a) and (b) and $a=2\xi$ for (c) and (d).} \label{fig:LDOSd+id}
\end{figure}
In the Fig.\ref{fig:LDOSd+is} we show the profile of zero-energy
LDOS in the case when the vortex is placed at a distance of
$a=2\xi$ from a flat boundary of a $d+is$-wave superconductor
characterized by a gap function in momentum space given by
Eq.(\ref{GapD+iS}). In this section, we use the notation $\xi=\hbar
v_F/\Delta_0$.

One can see that for one sign of vorticity
 the surface LDOS shows two
peaks which are symmetric with respect to the vortex position. As
we have shown above, the large peaks in surface LDOS appear when
the energy coincides with the position of bound state level. For
a different sign of the vorticity, there are no surface states at the Fermi
level and the LDOS along the surface is a flat function. A
non-zero level of LDOS in this case is provided by inelastic
scattering which leads to the smearing the QP energy levels.
Applying a local Doppler shift approach, which holds if the
distance from vortex to surface is rather large ($a\gg \xi$) one can
interpret the results shown in Fig.(\ref{fig:LDOSd+is}).

The coordinates $y^*$ of surface LDOS peaks can be estimated from
the relation $v_{sy}=\Delta_s/p_F$, where
 $v_{sy}=(M\hbar/m)a/(y^2+a^2)$ is a projection on the surface plane of superfluid
  velocity generated by the vortex with vorticity $M$ and image
  antivortex. It can be seen that for
  $a>\xi(\Delta_0/|\Delta_s|)$ the peak is situated at $y^*=0$
  i.e. at the surface point nearest to the vortex. Otherwise, we obtain $y^*=\pm a
  \sqrt{1-(a/\xi)(|\Delta_s|/\Delta_0)}$. Comparing this
  estimation with the numerical results in Fig.\ref{fig:LDOSd+is}, one
  observes a minor difference. For example, it follows from the
  estimation that the LDOS peaks should be positioned at $y^*=\pm 2.4\xi$
  for $\Delta_s=0.2\Delta_0$, but in Fig.\ref{fig:LDOSd+is} they
  are located at $y^*=\pm 2.0\xi$.
 This discrepancy can be attributed to the
  complex shift of the energy $\varepsilon \to \varepsilon+i\delta$
 due to the effective scattering
  parameter $\delta=0.1\Delta_0 $ which was used in the numerical
  calculations.
  If we increase the distance from vortex to surface $a$, the LDOS peaks
   will merge when $a>\xi (\Delta_0/|\Delta_s|)$ (see
Fig.\ref{fig:LDOSd+is}).

In Fig.\ref{fig:LDOSd+id}, we show the LDOS profile modulated
by a vortex placed at a distance of $a=2\xi$ from a flat
boundary of $d+id$ superconductor. The structure of the gap function
was chosen in the form (\ref{GapD+iD}). Applying the approach
based on the local Doppler shift we obtain the similar expression
for the coordinates of the peaks of surface LDOS: $y^*=\pm a
  \sqrt{1-(a/\xi)(|\Delta_d|/\Delta_0)}$. For the particular values of parameters
   $\Delta_d=0.2\Delta_0$ and $a=2\xi$ this estimation yields $y^*=\pm
   2.4\xi$,  which is much less than
 obtained from numerical plot in Fig.\ref{fig:LDOSd+id} $y^*\approx\pm 4\xi$.
This discrepancy can also be attributed to the effect of inelastic
scattering, which appears to have a larger effect in $d+id$- wave
case than in discussed above $d+is$-wave case.

\begin{figure}[h!]
\centering \resizebox{0.45\textwidth}{!}{
\includegraphics{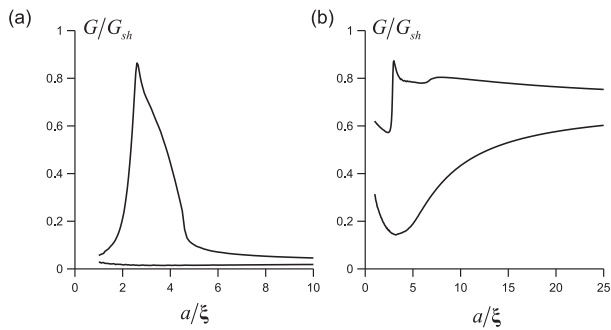}}
\caption{ Plots of the vortex-induced conductance
 in case of (a) chiral $d+is$ superconductor for
 $\Delta_s=0.2\Delta_0$ and (b) $d+id$ superconductor for
  $\Delta_d=0.2\Delta_0$. The strength of interface barrier is $Z=5$.
 Different curves on each plots correspond to different vortex orientations
 with respect to the $z$- axis. } \label{fig:Cond_Dwave}
\end{figure}

A numerical plot of the N/S junction conductance as a function of
distance from vortex to surface is shown in
Fig.\ref{fig:Cond_Dwave} for the $d+is$ and $d+id$ cases. The
conductance is normalized to the Sharvin conductance
$G_{sh}=k_FL_y/\pi$. Comparing Fig.\ref{fig:Cond_Dwave}a and
Fig.\ref{fig:Cond_Dwave}b one can see that the conductance
behaviour is qualitatively different for $s$ and $d$- wave
symmetry of the additional gap function component. For $d+is$-
wave, the conductance has a sharp peak for one vortex orientation
(upper curve in Fig.\ref{fig:Cond_Dwave}a) and it is a flat
function of $a$ for another vortex orientation (lower curve in
Fig.\ref{fig:Cond_Dwave}a). The origin of the conductance
enhancement is a formation of Andreev bound states at the Fermi
level which are localized near the superconducting surface. As was
discussed in the Introduction (see the Fig. \ref{SpectrumTr}b),
the zero-energy Andreev bound states can appear only for a certain
direction of superfluid velocity flowing along the superconducting
surface and if the value of the superfluid velocity is larger than
a critical value $|v_{sy}|>|\Delta_s|/(\hbar k_F)$. For a high
interface barrier $Z\gg 1$ applying an approximate  analytical
expression (\ref{VortexConductance}) we find that a sharp increase
of conductance in Fig.\ref{fig:Cond_Dwave}a can be described by
the following expression
 $$
 G/G_{sh}=\frac{16\pi}{3Z^2}\frac{\xi}{L_y}\frac{\Delta_0}{|\Delta_s|}\left(1-\frac{a}{a^*}\right)^{3/2}
 +\lambda Z^{-4},
 $$ where
$a^*=\xi (\Delta_0/|\Delta_s|)$ and $\lambda\sim 1$. Otherwise, if
$a>a^*$ the conductance is much smaller since $Z\gg 1$:
 $$ G/G_{sh}\approx \left(\frac{\Delta_0}{\Delta_s}\right)^2
\frac{4}{3Z^4}.$$ When the distance $a$ is decreased further, the
conductance is suppressed (see Fig.\ref{fig:Cond_Dwave}a, upper
curve). The decrease of conductance can be attributed to the gap
at the Fermi level which appears due to the interaction of vortex
and surface states in a similar way as for the $p+ip$- wave case
discussed in the previous section.

In a $d+id$-wave superconductor, zero-energy Andreev bound states may
exist even in the absence of vortex. An asymptotic value of the
conductance $G_0$ at $a\gg\xi$ can be obtained using the
expression (\ref{VortexConductance}) as follows
 $$
 G_0/G_{sh}=\left(\frac{\Delta_0}{|\Delta_d|}\right)\frac{\pi}{2\sqrt{2}Z^2}.
 $$
 When the vortex approaches the superconducting surface, the conductance is
 either suppressed (lower curve in Fig.\ref{fig:Cond_Dwave}b) or
 slightly enhanced (upper curve in Fig.\ref{fig:Cond_Dwave}b).
 This behaviour can be understood by again using the Eqs.(\ref{VortexConductance})
 with the Doppler shifted spectrum of Andreev bound states
 (\ref{ABS-D+iD}). The decrease (increase) of conductance corresponds to the
 transformation of spectrum shown qualitatively in
 Fig.\ref{SpectrumTr}c
 by dash (dash-dotted) lines. It is possible to obtain an analytical expression
 for the vortex-induced conductance shift at $a\gg\xi$ in the
 following form:
 $$
 \delta G/G_{sh}=\pm
 \frac{\pi}{2Z^2}\left(\frac{\Delta_0}{\Delta_d}\right)^2\frac{\xi}{L_y}\arctan\left(\frac{L_y}{2a}\right),
 $$
 where the upper and lover signs correspond to the different
 vortex orientations.
 As the vortex approaches the surface further, there appears an extremum of the conductance. Such
 behaviour can be explained by a conductance enhancement due to
 the tunneling of QP into the vortex core states, discussed in
 Ref.\onlinecite{Silaev}.
A sharp decrease of the upper curve in Fig.\ref{fig:Cond_Dwave}b
can be attributed to the opening of an energy gap at the Fermi
level due to the interaction of vortex and surface states.

\section{Summary}\label{sec:summary}

 In summary, we have investigated how the tunneling conductance and the local
density of states (LDOS) in superconductors are affected by the
influence of an external magnetic field when the superconducting
order parameter (OP) breaks time-reversal symmetry (TRS). This is
directly relevant for both Sr$_2$RuO$_4$, where chiral $p+ip$-wave
pairing is believed to be realized, and for the high-$T_c$
cuprates, where a $d+is$- or $d+id$-wave OP has been suggested to
exist near surfaces. In addition to breaking TRS, all of these OPs
feature surface bound zero-energy states at surfaces under
appropriate circumstances (e.g. a dominant $d$-wave OP in the
$d+is$-wave case).
\par
We have shown how the Doppler-shift conspires with an interaction
of vortex and surface states to produce a considerable qualitative
modification of both the tunneling conductance and the LDOS. When
the vortex is located at distances well above a coherence length
$\xi$ from the surface, the Doppler-shift produces an
enhancement or suppression of the LDOS depending on the relative
sign of the vorticity and the chirality of the superconducting OP.
This effect may be directly probed by first applying an external
magnetic field in a direction while measuring the LDOS and then
reversing the field direction and measuring again. When the vortex
is located very close to the surface (a distance of order $\xi$ or
smaller), there is an overlap between the vortex and surface
states which effectively cause a dramatic change in the tunneling
conductance and LDOS. This effect is also sensitive to the
relative sign of the vorticity and the chirality of the
superconducting OP. The overlap between these two sets of states
results in either a strongly enhanced or suppressed tunneling
conductance/LDOS at zero bias voltage/zero energy.
\par
We have demonstrated the aforementioned effects both
qualitatively and quantitatively for $p+ip$-, $d+is$-, and
$d+id$-wave symmetries. Experimentally, the distance from the surface to the closest vortex can be
altered by modifying the field strength. All of our predictions
should be possible to test experimentally with present-day techniques.

\section{acknowledgments}
 M.S. is grateful to Alexander S. Mel'nikov for
numerous stimulating discussions. This work was supported, in
part, by Russian Foundation for Basic Research, by Program
''Quantum Macrophysics'' of RAS, and by Russian Science Support
and ''Dynasty'' Foundations. J.L. and A.S. were supported by the
Norwegian Research Council Grant Nos. 158518/431, 158547/431,
(NANOMAT), and 167498/V30 (STORFORSK). T.Y. acknowledges support
by JSPS.


\begin{thebibliography}{99}

\bibitem{Pwave}
Y. Maeno, H. Hashimoto, K. Yoshida, S. Nishizaki, T. Fujita, J. G.
Bednorz, and F. Lichtenberg, Nature (London) {\bf 372}, 532
(1994); K. Ishida, H. Mukuda, Y. Kitaoka, K. Asayama, Z. Q. Mao,
Y. Mori, and Y. Maeno, Nature (London) {\bf 396}, 658 (1998); G.
M. Luke, Y. Fudamoto, K. M. Kojima, M. I. Larkin, J. Merrin, B.
Nachumi, Y. J. Uemura, Y.Maeno, Z. Q. Mao, Y. Mori, H. Nakamura,
and M. Sigrist, Nature (London) {\bf 394}, 558 (1998); A. P.
Mackenzie and Y. Maeno, Rev. Mod. Phys. {\bf 75}, 657 (2003);  K.
D. Nelson, Z. Q. Mao, Y. Maeno, and Y. Liu, Science {\bf 306},
1151 (2004); Y. Asano, Y. Tanaka, M. Sigrist, and S. Kashiwaya,
Phys. Rev. B {\bf 67}, 184505 (2003); Phys. Rev. B 71, 214501
(2005).

\bibitem{Buchholtz1}
L. J. Buchholtz, M. Palumbo, D. Rainer, and J. A. Sauls, J. Low
Temp. Phys. {\bf 101}, 1079 (1995).

\bibitem{Matsumoto1}
M. Matsumoto and H. Shiba, J. Phys. Soc. Jpn. {\bf 64}, 3384
(1995).

\bibitem{Sigrist}
K. Kuboki and M. Sigrist, J. Phys. Soc. Jpn. {\bf 65}, 361 (1996).

\bibitem{Sigrist_PTP}
M. Sigrist, Prog. Theor. Phys. {\bf 99}, 899 (1998).

\bibitem{Buchholtz2}
L. J. Buchholtz, M. Palumbo, D. Rainer, and J. A. Sauls, J. Low
Temp. Phys. {\bf 101}, 1099 (1995).

\bibitem{Matsumoto2}
M. Matsumoto and H. Shiba, J. Phys. Soc. Jpn. {\bf 64}, 4867
(1995).

\bibitem{Fogelstrom}
M. Fogelstrom, D. Rainer, and J. A. Sauls, Phys. Rev. Lett. {\bf
79}, 281 (1997).

\bibitem{Tanuma}
Y. Tanuma, Y. Tanaka, M. Ogata, and S. Kashiwaya, J. Phys. Soc.
Jpn. {\bf 67}, 1118  (1998); Phys. Rev. B {\bf 60}, 9817 (1999).

\bibitem{High-Tc}
M. Covington, M. Aprili, E. Paraoanu, L. H. Greene, F. Xu, J. Zhu,
and C. A. Mirkin, Phys. Rev. Lett. {\bf 79}, 277 (1997); A.
Biswas, P. Fournier, M. M. Qazilbash, V. N. Smolyaninova, H.
Balci, and R. L. Greene, Phys. Rev. Lett. {\bf 88}, 207004 (2002);
Y. Dagan and G. Deutscher, Phys. Rev. Lett. {\bf 87}, 177004
(2001). A. Sharoni, O. Millo, A. Kohen, Y. Dagan, R. Beck, G.
Deutscher, and G. Koren, Phys. Rev. B {\bf 65}, 134526 (2002); A.
Kohen, G. Leibovitch, and G. Deutscher, Phys. Rev. Lett. {\bf 90},
207005 (2003); M. Aprili, E. Badica, and L. H. Greene, Phys. Rev.
Lett. {\bf 83}, 4630 (1999); R. Krupke and G. Deutscher, Phys.
Rev. Lett. {\bf 83}, 4634 (1999).

\bibitem{FT}
M. Franz and Z. Tesanovic, Phys. Rev. Lett., {\bf 80}, 4763
(1998).

\bibitem{Hu}
C. R. Hu, Phys. Rev. Lett. {\bf 72}, 1526 (1994).

\bibitem{yang_prb_94} J. Yang and C.-R. Hu, Phys. Rev. B \textbf{50}, 16766 (1994).

\bibitem{Tanaka}
Y. Tanaka and S. Kashiwaya, Phys. Rev. Lett. {\bf 74}, 3451
(1995); S. Kashiwaya and Y. Tanaka, Rep. Prog. Phys.
{\bf 63} 1641 (2000).

\bibitem{Barash_Penetration}
L. Alff, S. Kleefisch, U. Schoop, M. Zittartz, T. Kemen, T. Bauch,
A. Marx, and R. Gross, Eur. Phys. J. B {\bf 5}, 423 (1998); H.
Walter, W. Prusseit, R. Semerad, H. Kinder, W. Assmann, H. Huber,
H. Burkhardt, D. Rainer, and J. A. Sauls, Phys. Rev. Lett. {\bf
80}, 3598 (1998); Yu. S. Barash, M. S. Kalenkov, and J.
Kurkija¨rvi, Phys. Rev. B {\bf 62}, 6665 (2000); A. Carrington, F.
Manzano, R. Prozorov, R. W. Giannetta, N. Kameda, and T. Tamegai,
Phys. Rev. Lett. {\bf 86}, 1074 (2001).

\bibitem{Barash_Josephson}
Y. Tanaka and S. Kashiwaya, Phys. Rev. B {\bf 53}, 11 957 (1996);
Yu. S. Barash, H. Burkhardt, and D. Rainer, Phys. Rev. Lett. {\bf
77}, 4070 (1996); R. A. Riedel and P. F. Bagwell, Phys. Rev. B
{\bf 57}, 6084 (1998); E. Il'ichev, M. Grajcar, R. Hlubina, R. P.
J. Ijsselsteijn, H. E. Hoenig, H. G. Meyer, A. Golubov, M. H. S.
Amin, A. M. Zagoskin, A. N. Omelyanchouk, and M. Yu. Kupriyanov,
Phys. Rev. Lett. {\bf 86}, 5369 (2001).

\bibitem{Barash_Chiral}
Yu. S. Barash, A. M. Bobkov, and M. Fogelstrom, Phys. Rev. B, {\bf
64}, 214503 (2000).



 \bibitem{wei_prl_98} J. Y. Wei, N.-C. Yeh, D. F. Garrigus, and M. Strasik, Phys. Rev. Lett. \textbf{81}, 2542 (1998).

 \bibitem{mao_prl_01} Z. Q. Mao, K. D. Nelson, R. Jin, Y. Liu, and Y. Maeno, Phys. Rev. Lett. \textbf{87}, 037003 (2001).


\bibitem{graser_prl_04} S. Graser, C. Iniotakis, T. Dahm, and N. Schopohl, Phys. Rev. Lett. \textbf{93}, 247001 (2004).

\bibitem{graser_prb_05} C. Iniotakis, S. Graser, T. Dahm, and N. Schopohl, Phys. Rev. B \textbf{71}, 214508 (2005).

\bibitem{yokoyama_prl_08} T. Yokoyama, C. Iniotakis, Y. Tanaka, and M. Sigrist, Phys. Rev. Lett. \textbf{100},
 177002 (2008).

\bibitem{Kopnin-Volovik-1996}
N.~B. Kopnin and G.~E. Volovik, JETP Lett. {\bf 64}, 690 (1996).


\bibitem{Maki}
Y. Morita, M. Kohmoto, and K. Maki, Phys. Rev. Lett., {\bf 78},
4841 (1997).

\bibitem{FTVM} M. Franz and Z. Tesanovic, Phys. Rev. Lett., {\bf 87}, 257003 (2000);
 O. Vafek, A. Melikyan, M. Franz, and Z. Tesanovic, Phys. Rev. B {\bf 63}, 134509 (2001);
   O. Vafek, A. Melikyan, and Z. Tesanovic, {\it ibid}  {\bf 64}, 224508 (2001);
{\it ibid}, {\bf 76}, 094509 (2007).

\bibitem{Melnikov_Dwave}
A.S. Mel'nikov, Phys. Rev. Lett., {\bf 86}, 4108 (2001).

\bibitem{Volovik}
 G. E. Volovik, JETP Lett., {\bf 70},
609 (1999).

\bibitem{Tinkham} M. Tinkham, {\it Introduction to
Superconductivity} (McGraw-Hill, New York, 1996), 2nd ed., Chap.
10.

\bibitem{Sigr}
J. Goryo and M. Sigrist, J. Phys. C, {\bf 12} L599 (2000).

\bibitem{CdGM}
C.~Caroli, P.~G. de Gennes, J.~Matricon, Phys. Lett. \textbf{9},
307 (1964).

\bibitem{Kopnin_BSrule}
N.~B. Kopnin, Phys.~Rev.~B {\bf 57}, 11775 (1998).


\bibitem{Mahan}
G.~D. Mahan, {\it Many-particle physics} (Plenum Press, New York,
1993), 2nd ed., Chap. 9.

\bibitem{BTK}
G.~E. Blonder, M.~Tinkham and T.~M. Klapwijk, Phys. Rev.~B
\textbf{25}, 4515 (1982).

\bibitem{Bardeen}
J.~Bardeen, R.~Kummel, A.~E. Jacobs and L.~Tewordt, Phys. Rev.
\textbf{187}, 556 (1969).

\bibitem{dahm_prb_02} T. Dahm, S. Graser, C. Iniotakis, and N. Schopohl, Phys. Rev. B \textbf{66}, 144515 (2002).

\bibitem{schopohl_prb_95} N. Schopohl and K. Maki, Phys. Rev. B \textbf{52}, 490 (1995); N. Schopohl, cond-mat/9804064.

\bibitem{covington_prl_97} M. Covington, M. Aprili, E. Paraoanu, L. H. Greene, F. Xu, J.
Zhu, and C. A. Mirkin, Phys. Rev. Lett. \textbf{79}, 277 (1997).

\bibitem{krupke_prl_00} R. Krupke and G. Deutscher, Phys. Rev. Lett. \textbf{83}, 4634 (2000).

\bibitem{sharoni_epl_01} A. Sharoni, G. Koren, and O. Millo, Europhys. Lett. \textbf{54}, 675 (2001).

\bibitem{Silaev} M.A. Silaev, Phys. Rev. B \textbf{77}, 014504 (2008).



\end{thebibliography}
\end{document}